# Efficient Spin-Orbit Torque Generation in Semiconducting WTe$_2$ with Hopping Transport

Cheng-Wei Peng,[1,a)] Wei-Bang Liao,[1,a)] Tian-Yue Chen,[1] and Chi-Feng Pai[1,2,b)]

[1]Department of materials science and engineering, National Taiwan University, Taipei 10617, Taiwan

[2]Center of Atomic initiative for new materials, National Taiwan University, Taipei 10617, Taiwan

[a)]These authors contributed equally to this work.

[b)]Author to whom correspondence should be addressed: Email: cfpai@ntu.edu.tw

**Abstract**

Spin-orbit torques (SOTs) from transition metal dichalcogenides systems (TMDs) in conjunction with ferromagnetic materials are recently attractive in spintronics for their versatile features. However, most of the previously studied crystalline TMDs are prepared by mechanical exfoliation, which limits their potentials for industrial applications. Here we show that amorphous WTe$_2$ heterostructures deposited by magnetron sputtering possess a sizable damping-like SOT efficiency $\xi_{\text{DL}}^{\text{WTe}_2} \approx 0.20$ and low damping constant $\alpha = 0.009 \pm 0.001$. Only an extremely low critical switching current density $J_\text{c} \approx 7.05 \times 10^9$ A/m$^2$ is required to achieve SOT-driven magnetization switching. The SOT efficiency is further proved to depend on the W and Te relative compositions in the co-sputtered W$_{100-x}$Te$_x$ samples, from which a sign change of $\xi_{\text{DL}}^{\text{WTe}_2}$ is observed. Besides, the electronic transport in amorphous WTe$_2$ is found to be semiconducting and is governed by a hopping mechanism. With the above advantages and rich tunability, amorphous and semiconducting WTe$_2$ serves as a unique SOT source for future spintronics applications.

## I. Introduction

With the advancement of spintronics and next generation magnetic random access memory (MRAM) technologies, spin-orbit torques (SOTs) originated from the spin-orbit interactions in various types of materials systems and magnetic heterostructures have been shown to be an effective mechanism to manipulate and switch the magnetization similar to or more efficient than the traditional spin-transfer torque (STT) approach.[1-4] When it comes to materials with sizable efficiency of generating spin current from charge current, the 5d transition metals like Pt,[1,2] Ta,[5] and W[6-9] are the typically chosen due to the strong spin-orbit coupling (SOC) therein, which can induce spin currents and SOTs through the spin Hall effect (SHE).[5,10,11]

More recently, topological insulators (TIs) such as BiSe,[12-15] BiSb,[16] and $(Bi_{0.5}Sb_{0.5})_2Te_3$[17] are reported to possess high SOT efficiency from spin momentum locking of topologically-protected surface states (TSSs). Interestingly, various non-epitaxial Bi-based chalcogenides without TSS have also been reported to show giant SOT efficiencies.[18-20] Another family of emergent materials, transition metal dichalcogenides (TMDs), have also gained lots of attention due to not only the strong SOC but also their unconventional SOT features from the lack of inversion symmetry.[21-32] These unique properties lead to the possibility of electric field control over SOT[24] and generation of out-of-plane damping-like (DL) SOTs.[22,23,25,28,30] Particularly, for exfoliated $WTe_2$, MacNeill *et al.* confirmed the existence of anomalous DL-SOT in $WTe_2$/Py through spin-torque ferromagnetic resonance (ST-FMR) measurements.[22] Li *et al.* observed an enhanced spin conductivity in $WTe_2$/Py originated from spin momentum locking.[26] Shi *et al.* reported an extremely large DL-SOT efficiency up to 0.5 for $WTe_2$/Py, which was attributed to a bulk origin.[27] Although all these studies suggest that $WTe_2$ is a promising SOT material with rich tunability, most of them require mechanical exfoliation of single-crystalline $WTe_2$, which is of great challenge to realize in mass production. It is therefore crucial to explore the possibilities of employing $WTe_2$ layers prepared by conventional materials growth approaches, such as magnetron sputter deposition, to see their spin transport properties as compared to the exfoliated cases.

In this work, we report the DL-SOT efficiencies from both stoichiometric WTe$_2$ and co-sputtered W$_{100-x}$Te$_x$-based magnetic heterostructures prepared by high vacuum sputter depositions. Both series of deposited multilayer stacks are amorphous. We first quantify the damping constant of WTe$_2$/CoFeB devices with in-plane magnetic anisotropy (IMA) through ST-FMR measurements,[33] from which the damping constant is determined to be $\alpha_{\text{WTe}_2/\text{CoFeB}} = 0.009 \pm 0.001$, smaller than those from the transition metal-based control samples, namely $\alpha_{\text{W/CoFeB}} \approx 0.014$ and $\alpha_{\text{Pt/CoFeB}} \approx 0.033$. The largest DL-SOT efficiency from the stoichiometric WTe$_2$/CoTb devices with perpendicular magnetic anisotropy (PMA) is further estimated to be $\xi_{\text{DL}}^{\text{WTe}_2} \approx 0.20$ by current-induced hysteresis loop shift measurement.[34] Current-induced magnetization switching at a low critical switching current density $J_c \approx 7.05 \times 10^9$ A/m$^2$ further confirms the sizable DL-SOT efficiency. By studying the co-sputtered W$_{100-x}$Te$_x$ samples, we find that $\xi_{\text{DL}}^{\text{W}_{100-x}\text{Te}_x}$ is strongly composition dependent in terms of both magnitude and sign, which can explain the huge discrepancies of observed $\xi_{\text{DL}}$ in several recent reports.[35,36] Additionally, the distinct difference between the layer thickness dependence of $\xi_{\text{DL}}^{\text{WTe}_2}$ and $\xi_{\text{DL}}^{\text{W}_{100-x}\text{Te}_x}$ suggests that both bulk and interfacial origins of SOTs are possible to be generated by sputter-deposited chalcogenides. These unconventional features together with its low $\alpha$ and sizable $\xi_{\text{DL}}$, make amorphous WTe$_2$-based heterostructure an intriguing system for novel SOT applications.

## II. Materials and Methods

The samples in this work are all prepared by magnetron sputtering and deposited onto thermally oxidized silicon substrates with base pressure $\sim 3 \times 10^{-8}$ Torr. We use dc sputtering for metallic targets under 3 mTorr of Ar working pressure, while MgO target is rf sputtered under 10 mTorr of Ar working pressure. Three series of samples are prepared, namely WTe$_2$(10)/CoFeB(5)/MgO(1)/Ta(1) for ST-FMR measurements, WTe$_2$(t)/CoTb(6)/Ta(2) for hysteresis loop shift and current-induced magnetization switching measurements, and co-sputtered W$_{100-x}$Te$_x$ (10)/CoTb(4.5)/Ta(2) (units in nanometers) for further comparisons. The WTe$_2$ layer is deposited directly from a stoichiometric WTe$_2$ target, while the CoFeB layer is sputtered from a Co$_{40}$Fe$_{40}$B$_{20}$ target. The Co$_{55}$Tb$_{45}$ (CoTb) layer and W$_{100-x}$Te$_x$ layer are deposited by co-sputtering Co and Tb targets as well as W and Te targets, respectively. Note that WTe$_2$ layer is grown at a substrate temperature of 300°C to ensure the uniformity.[37] The thicknesses of each layer are checked by atomic force microscopy (AFM). Electron probe X-ray microanalyzer (EPMA) is employed to confirm the atomic ratios of the sputtered WTe$_2$ and co-sputtered W$_{100-x}$Te$_x$ thin films. For further electrical measurements, the films are patterned into microstrip with lateral dimensions of 40 μm×100 μm for ST-FMR measurement and Hall bar devices with lateral dimensions of 10 μm×60 μm for hysteresis loop shift and current-induced magnetization switching measurements through standard photolithography and followed by Ar ion-milling.

## III. Results

### A. Materials characterizations

The structural properties of the sputtered WTe$_2$ films are first examined by cross-sectional field-emission transmission electron microscopy (FE-TEM, JEOL 2010F). The TEM sample is prepared by a lift-out technique with SEIKO SMI-3050SE focused ion beam (FIB). As shown in Fig. 1(a), the sputtered WTe$_2$ in a representative WTe$_2$(10)/CoFeB(5)/MgO(1)/Ta(1) sample is amorphous. Figure 1(b) shows the normalized secondary ion mass spectrometer (SIMS) signals from the same thin film during Ar ion-milling process. The peaks of Mg, Co, W, and Te correspond to the signals from the MgO layer, the CoFeB layer, and the WTe$_2$ layer being etched, respectively. According to the EPMA analysis, the atomic ratio between W and Te are 33% and 67% in our sputtered WTe$_2$ films, therefore we keep the stoichiometric notation of WTe$_2$ to represent samples deposited from the WTe$_2$ target (see the supplementary material S1). The resistivities of each material are measured by four-point probe measurement, among which $\rho_{\text{WTe}_2} \approx 1870$ μΩ-cm (see the supplementary material S2). We further examine the electrical transport properties by measuring the temperature dependence of resistivity from 323 K to 423 K, which suggests our sputtered WTe$_2$ and W$_{100-x}$Te$_x$ films are semiconducting rather than metallic (see the supplementary material S3).

We further examine the electrical transport properties by measuring the temperature dependence of resistivity from 323 K to 423 K. Most of the previous works claimed that the textured WTe$_2$ is one kind of Weyl semimetal,[26,27] while our sample is predominately amorphous and therefore should not fall into this category. According to the electrical transport measurement shown in supplementary material S3, the sputtered amorphous WTe$_2$ is semiconducting and the temperature dependent resistivity can be well described by small polaron hopping (SPH) model,[38] where the thermally-activated hopping polarons from electron-phonon interaction dominate the electrical transport.

## B. ST-FMR measurement

We then investigate the spin transport properties of WTe$_2$(10)/CoFeB(5)/MgO(1)/Ta(1) with ST-FMR. CoFeB is chosen to be the ferromagnetic (FM) layer because it is widely used in contemporary MTJs.[1,39,40] MgO(1)/Ta(1) layers are employed to prevent the layers beneath from oxidation. In a typical ST-FMR measurement, a rf current is applied, generating a rf spin current from the SOC buffer layer (WTe$_2$) to induce the precessional magnetization of the FM layer, which leads to an oscillation of the resistance owing to the anisotropic magnetoresistance. With the mixing of oscillating resistance and rf current, we can measure a rectified dc voltage $V_{\text{mix}}$ of the device through a bias tee, as illustrated in Fig. 2(a). A rf current is generated from the signal generator with the amplitude modulated at 500 Hz, which allows the use of lock-in amplifier for $V_{mix}$ detection. A sweeping in-plane field $H_{ext}$ is applied at a fixed angle of 135º between $H_{ext}$ and the current channel. Typical field-swept ST-FMR spectra of a WTe$_2$(10)/CoFeB(5) microstrip under frequencies of $f = 7$ to 12 GHz are shown in Fig. 2(b). The spectrum of $f = 10$ GHz is further scrutinized in Fig. 2(c), in which the data points of $V_{mix}$ can be fitted with[33]

$$V_{mix} = S\frac{\Delta^2}{\Delta^2 + (H_{ext} - H_0)^2} + A\frac{\Delta(H_{ext} - H_0)}{\Delta^2 + (H_{ext} - H_0)^2}, \quad (1)$$

where $S$, $A$, $\Delta$, and $H_0$ denote symmetric Lorentzian coefficient, anti-symmetric Lorentzian coefficient, linewidth, and resonant field, respectively. The symmetric part of $V_{mix}$ originates from the in-plane DL torque ($\tau_{\parallel}$), while the anti-symmetric part comes from the out-of-plane field-like (FL) torque ($\tau_{\perp}$) and the Oersted field. The torque ratio is further evaluated by[24,33]

$$\frac{\tau_{\parallel}}{\tau_{\perp}} = \frac{S}{A}\sqrt{1 + \frac{4\pi M_{eff}}{H_0}}, \quad (2)$$

where $4\pi M_{eff} \approx 11.7$ kOe is the demagnetization field extracted from the fitting of Kittel formula (see the supplementary material S4). Note that the contribution from the Oersted field is negligible in our system. For the WTe$_2$(10)/CoFeB(5) device, the torque ratio is $2.75 \pm 0.54$, which indicates a considerable DL torque contribution. However, in ST-FMR measurement, other factors like the FM thicknesses dependent FL-SOT[41,42] and spin-pumping effect[43,44] might plague the estimation of $\xi_{DL}^{WTe_2}$ from torque ratio. Therefore, we choose to quantify $\xi_{DL}^{WTe_2}$ via other approaches, which will be addressed in the following section.

Nevertheless, ST-FMR provides us other valuable information, such as damping constant

$\alpha$, which is one of the key figures of merit to realize efficient current-induced switching.[45,46] $\alpha$ can be estimated from $\Delta = \Delta_0 + \frac{2\pi\alpha}{\gamma}f$, where $\Delta_0$ is the linewidth from inhomogeneous broadening, and $\gamma$ is the gyromagnetic ratio. Figure 2(d) shows the linewidths as functions of frequencies of WTe$_2$, Pt and W-based magnetic heterostructures. Surprisingly, the WTe$_2$(10)/CoFeB(5) device shows $\alpha_{\text{WTe}_2/\text{CoFeB}} = 0.009 \pm 0.001$, which is smaller than those obtained from control samples with classical spin Hall transition metals, namely Pt(6)/CoFeB(5) ($\alpha_{\text{Pt/CoFeB}} = 0.033 \pm 0.003$) and W(4)/CoFeB(5) ($\alpha_{\text{W/CoFeB}} = 0.014 \pm 0.002$). Note that $\alpha_{\text{WTe}_2/\text{CoFeB}}$ from our sputtered WTe$_2$/CoFeB is comparable to those observed from exfoliated WTe$_2$/Py heterostructures.[22,27]

## C. Hysteresis loop shift measurement

Next, we employ hysteresis loop shift measurement[34] to quantify the DL-SOT efficiency $\xi_{\text{DL}}^{\text{WTe}_2}$ of WTe$_2$/FM heterostructures, with FM being ferrimagnetic CoTb alloy with PMA. Ferrimagnets are widely used as SOT detector due to their robust bulk PMA on diverse materials with SOC.[13,47,48] The setup for loop shift measurement on a WTe$_2$(10)/CoTb(6)/Ta(2) Hall bar device with suitable PMA ($H_c \approx 60$ Oe) is depicted in Fig. 3(a). A dc current ($I_{\text{dc}}$) is applied along the current channel with a parallel in-plane bias field $H_x$. With $H_x$ overcoming the effective field from interfacial Dzyaloshinskii-Moriya interaction (DMI) $H_{\text{DMI}}$ and realigning the domain wall moments, the current-induced effective field from DL-SOT will assist domain wall propagation and domain expansion.[49,50] When $|H_x| \geq |H_{\text{DMI}}|$, the current-induced DL-SOT from WTe$_2$ can fully act on the domain wall moments of CoTb and it can be seen as an out-of-plane effective field $H_z^{\text{eff}}$ causing the hysteresis loop to shift. This can be monitored via anomalous Hall voltage measurements, as shown in Fig. 3(b). By obtaining the slope of $H_z^{\text{eff}}$ vs $I_{\text{dc}}$, as shown in Fig. 3(c), the DL-SOT efficiency ($\xi_{\text{DL}}$) can be estimated through[34]

$$\xi_{DL} = \frac{2e}{\hbar}\left(\frac{2}{\pi}\right)\mu_0 M_s t_{CoTb} w t_{WTe_2} \left(\frac{\rho_{CoTb}t_{WTe_2} + \rho_{WTe_2}t_{CoTb}}{\rho_{CoTb}t_{WTe_2}}\right)\left(\frac{H_z^{eff}}{I_{dc}}\right), \qquad (3)$$

where $M_s \approx 48$ emu/cm$^3$ is the saturation magnetization of the CoTb layer (characterized by vibrating sample magnetometer (VSM)), $w = 10$ μm is the width of the Hall bar device, and $\rho$ represents the layer resistivity. The estimated DL-SOT efficiency of the sputtered WTe$_2$(10)/CoTb(6) heterostructure is $\xi_{DL}^{WTe_2} \approx 0.20$, which is comparable with the exfoliated crystalline WTe$_2$/FM.[27,31] It is worth noting that this sizable $\xi_{DL}^{WTe_2}$ is in sharp contrast to a recent work by Fan *et al.*, in which a relatively small $\xi_{DL} \approx -0.0021$ but a large $\xi_{FL} \approx 0.057$ are reported for the sputtered W$_x$Te$_{2-x}$.[35] Figure 3(d) shows $H_z^{eff}/I_{dc}$ ($\propto \xi_{DL}$) as a function of $H_x$ for WTe$_2$/CoTb and W/CoTb (control sample) devices. It can be observed that the signs of $H_z^{eff}/I_{dc}$ for the two devices are opposite ($\xi_{DL}^{WTe_2} > 0$ and $\xi_{DL}^{W} < 0$), which is consistent with the exfoliated WTe$_2$ case.[22,27,31] We emphasize that both the DL effective field ($H_z^{eff}/I_{dc}$) and the estimated DL-SOT efficiency of WTe$_2$-based device ($\xi_{DL}^{WTe_2} \approx 0.20$) are greater than those of the W-based control sample ($\xi_{DL}^{W} \approx -0.04$) (see the supplementary material S5). It is also interesting to note that $|H_{DMI}^{WTe_2}| \gg |H_{DMI}^{W}|$, which suggests that the WTe$_2$-based device might have a stronger interfacial SOC than the W-based case.

### D. Current-induced magnetization switching

We further demonstrate current-induced SOT-driven magnetization switching in WTe$_2$(10)/CoTb(6)/Ta(2) magnetic heterostructures. Pulsed currents are injected into the current channel of Hall bar devices, and again, an $H_x$ is applied to overcome $H_{DMI}$ such that the deterministic magnetization switching can be achieved. As shown in Fig. 4(a), the switching polarities depend on the applied $H_x$, which is consistent with the SOT-driven switching mechanism.[2,5] More importantly, the critical switching current density $J_c$ of WTe$_2$/CoTb structure is further demonstrated to be $J_c \approx 7.05 \times 10^9$ A/m$^2$, which is much lower than those from the 5d transition metal-based heterostructures ($J_c \approx 10^{11}$ A/m$^2$)[1,5,6] and comparable to the TI cases.[13-16] Note that this value is of the same order as the exfoliated 80 nm-thick-WTe$_2$ case reported by Shi. S et al. ($J_c \approx 3 \times 10^9$ A/m$^2$),[27] and smaller than the sputtered WTe$_x$ case reported by Li et al. ($J_c \approx 1.5 \times 10^{10}$ A/m$^2$).[36] In addition to $J_c$, the SOC layer thickness also needs to be taken into account to estimate the overall switching performance. We evaluate the ideal critical switching current through $I_c^{ideal} = J_c (t \cdot w)$, where $t$ is the required SOC buffer layer thickness and $w$ is the ideal current channel width assumed to be 100 nm. Figure 4(b) compares the reported $J_c$ and estimated $I_c^{ideal}$ from several TIs and WTe$_x$-based magnetic heterostructures (see the supplementary material S6 for parameter details). Both benchmark results indicate that amorphous sputtered WTe$_2$ is a competitive SOT source in the emergent chalcogenide materials category.

**E. Co-sputtered W$_{100-x}$Te$_x$ devices**

Lastly, we turn our focus on the possible origins of SOT in the amorphous WTe$_2$ heterostructures. Due to the lack of crystalline phase (and therefore no TSS) in our sample, we speculate that the SOC should have a bulk origin and the concentration of constituent elements will strongly affect $\xi_{DL}$. Figure 5(a) shows $\xi_{DL}$ obtained from a series of co-sputtered samples, $W_{100-x}Te_x(10)/CoTb(4.5)/Ta(2)$. It is found that $|\xi_{DL}|$ gradually increases and reaches a maximum at 18% of Te doping, where $\xi_{DL}^{W_{82}Te_{18}} \approx -0.09$. Further increasing the Te concentration will lead to a decrease of $|\xi_{DL}|$. Note that $\xi_{DL}$ changes sign from co-sputtered $W_{100-x}Te_x$/CoTb (negative) to $WTe_2$/CoTb (positive). This sign-change suggests that the governing SOC behind the W-rich and the Te-rich regimes could be different. Although there is few work discussing the SOC or SOT from pure Te, previous studies have pointed out that elements with *p*-orbitals are able to possess strong SOCs.[51-53] A recent first principles calculation also indicates the spin Hall conductivity (SHC) is sensitive to the Fermi energy of $WTe_2$, where the sign of SHC could reverse in a narrow energy window.[54]

To understand the possible difference in SOT generation from $WTe_2$ and $W_{100-x}Te_x$, we further examine the SOC layer thickness dependence of $|\xi_{DL}|$ from both type of samples. For the co-sputtered case, we use $W_{82}Te_{18}$/CoTb as the representative sample due to its maximized efficiency. As summarized in Fig. 5(b), the co-sputtered $W_{82}Te_{18}$ and the stoichiometric $WTe_2$ cases have a drastically difference in SOC layer dependence. For the co-sputtered case, the thickness dependence of efficiency can be well fitted by the spin diffusion model $\left|\xi_{DL}\left(t_{W_{82}Te_{18}}\right)\right| = \left|\xi_{DL}^{W_{82}Te_{18}}\right|\left[1 - \text{sech}\left(t_{W_{82}Te_{18}} / \lambda_s^{W_{82}Te_{18}}\right)\right]$[33], with $\left|\xi_{DL}^{W_{82}Te_{18}}\right| \approx 0.09$ and spin diffusion length $\lambda_s^{W_{82}Te_{18}} \approx 4.1$ nm, which suggests that the SOC within $W_{82}Te_{18}$ is of bulk origin (such as the SHE). This is perhaps due to a dominating contribution of the SHE from W in the W-rich regime. In contrast, for $WTe_2$, $\left|\xi_{DL}^{WTe_2}\right|$ decays with increasing the stoichiometric layer thickness, which indicates the possible existence of interfacial contributions of SOC besides the bulk effect.[14,29] We speculate that interfacial effects could play crucial roles on the SOT generation in the Te-rich (stoichiometric) regime, even if the TMD layer is amorphous. Further studies are required to elucidate the SOC and SOT generation mechanisms in these amorphous TMD systems.

## IV. Conclusions

In summary, we use conventional magnetron sputtering to deposit WTe$_2$ instead of mechanical exfoliation, where the structure is confirmed to be amorphous by TEM imaging. From the ST-FMR measurements, we obtain a sizable DL-SOT response with a low damping constant in WTe$_2$/CoFeB heterostructure ($\alpha_{\text{WTe}_2/\text{CoFeB}} \approx 0.01$). Through hysteresis loop shift measurements, the largest DL-SOT efficiency of WTe$_2$/CoTb heterostructure is characterized to be $\xi_{\text{DL}}^{\text{WTe}_2} \approx 0.20$. This large charge-to-spin conversion efficiency is further verified by current-induced magnetization switching with a low critical switching current density of $J_c \approx 7.05 \times 10^9$ A/m$^2$. By comparing the stoichiometric WTe$_2$ to the co-sputtered W$_{100-x}$Te$_x$-based devices, the sign and magnitude of $\xi_{\text{DL}}$ is found to be strongly composition dependent. Moreover, not only bulk SHE but also other interfacial SOC may contribute to the SOTs from WTe$_2$. Our observation therefore suggests that the relative composition between W and Te is the key factor, rather than the crystallinity, to affect the resulting spin transport properties, and the transport mechanism in amorphous sputtered WTe$_2$ can be described by the SPH model. With the features of wide-range tunable $\xi_{\text{DL}}$, low $\alpha$, and low $J_c$, amorphous sputtered WTe$_2$ can serve as a potential SOT source material in the semiconducting regime.

**Supplementary Material**

See the supplementary material associated with this article.

**Acknowledgements**

We thank Tsao-Chi Chuang and Hsia-Ling Liang for the support on AFM and VSM measurements. We also thank Mr. Chung-Yuan Kao of Ministry of Science and Technology (National Taiwan University) for the assistance in EPMA, and thank Ms. Chia-Ying Chien of Ministry of Science and Technology (National Taiwan University) for the assistance on focused ion beam (FIB). This work is supported by the Ministry of Science and Technology of Taiwan (MOST) under grant No. MOST-109-2636-M-002-006 and by the Center of Atomic Initiative for New Materials (AI-Mat) and the Advanced Research Center of Green Materials Science and Technology, National Taiwan University from the Featured Areas Research Center Program within the framework of the Higher Education Sprout Project by the Ministry of Education (MOE) in Taiwan under grant No. NTU-109L9008. The authors declare no competing financial interests.

**Author Contributions**

Cheng-Wei Peng prepared the IMA samples and performed ST-FMR measurement. Wei-Bang Liao prepared the PMA samples, performed hysteresis loop shift measurement and current-induced magnetization switching. Tian-Yue Chen conceived the experimental protocols and performed the analysis on the experimental data with Cheng-Wei Peng and Wei-Bang Liao. Chi-Feng Pai proposed and supervised the study.

**Data availability**

The data that support the findings of this study are available from the corresponding author upon reasonable request.

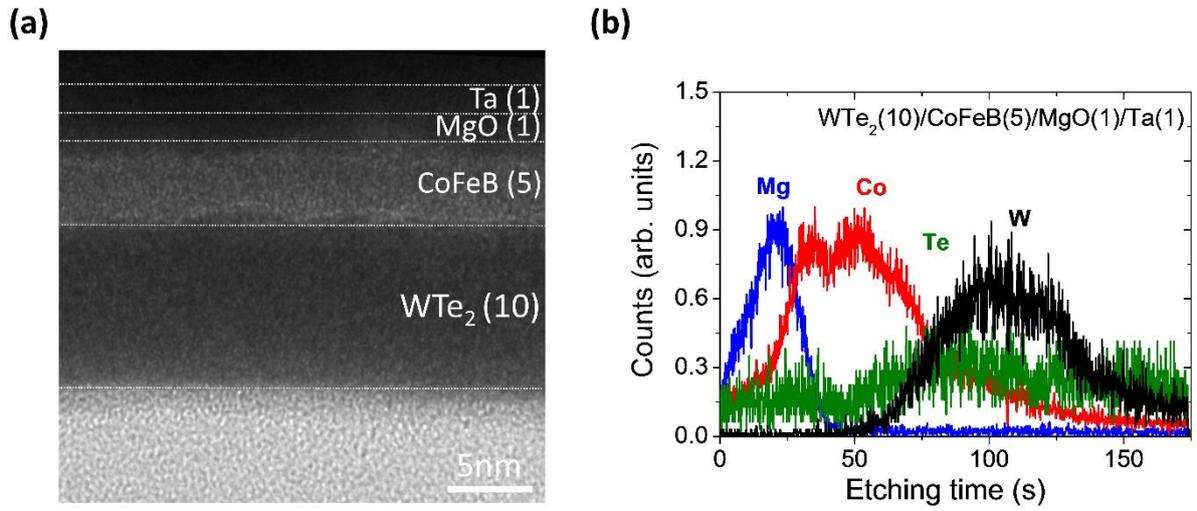

**FIG. 1.** Structural characterization of WTe$_2$-based heterostructure. (a) Cross-sectional TEM image and (b) SIMS signals with respect to the etching (Ar ion-milling) time of a representative WTe$_2$(10)/CoFeB(5)/MgO(1)/Ta(1) heterostructure.

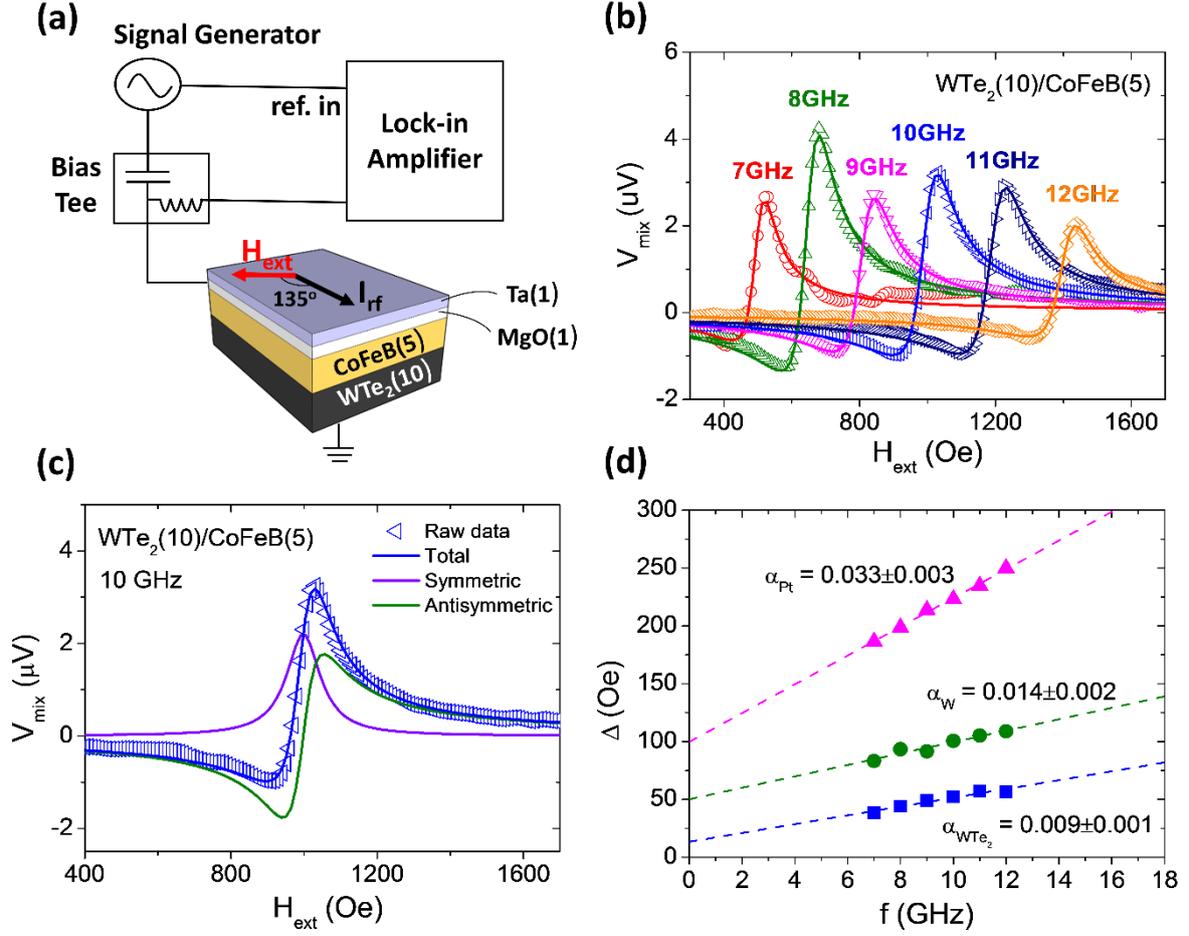

**FIG. 2.** ST-FMR measurement of WTe$_2$(10)/CoFeB(5)/MgO(1)/Ta(1) device. (a) Illustration of ST-FMR measurements setup. (b) The ST-FMR spectra of a WTe$_2$(10)/CoFeB(5)/MgO(1)/Ta(1) microstrip for $f = 7$ to 12 GHz. (c) Representative ST-FMR result of WTe$_2$(10)/CoFeB(5)/MgO(1)/Ta(1) at $f = 10$ GHz. The raw data is fitted by the sum of a symmetric and an antisymmetric Lorentzian. (d) $f$ dependence of linewidth $\Delta$ for Pt(6)/CoFeB(5)/MgO(1)/Ta(1), W(4)/CoFeB(5)/MgO(1)/Ta(1) and WTe$_2$(10)/CoFeB(5)/MgO(1)/Ta(1) structures.

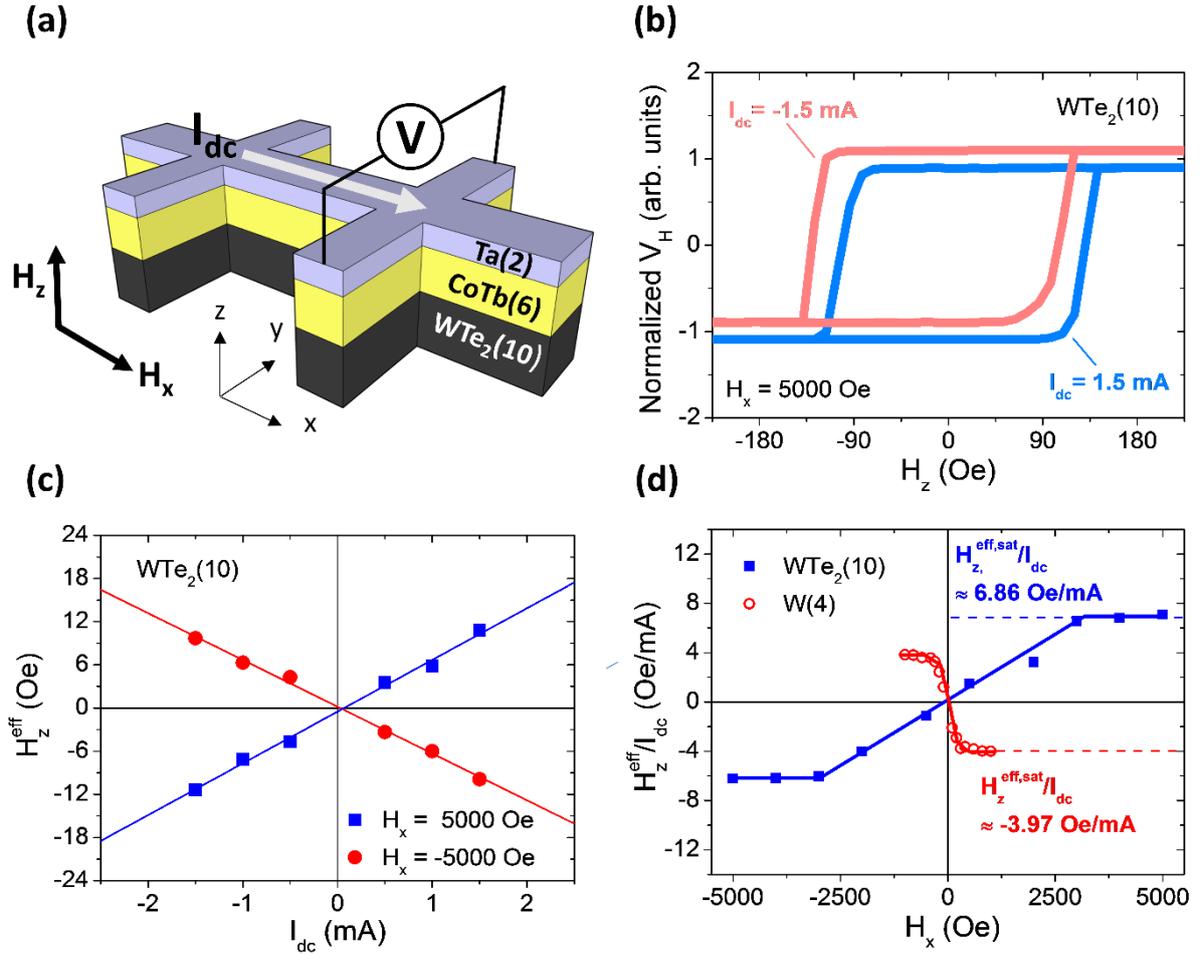

**FIG. 3.** Hysteresis loop shift measurement of WTe$_2$(10)/CoTb(6)/Ta(2) device. (a) Illustration of a Hall bar device for hysteresis loop shift measurement. (b) Representative shifted hysteresis loops of a WTe$_2$(10)/CoTb(6)/Ta(2) device with $I_{dc} = \pm 1.5$ mA and $H_x = 5000$ Oe. (c) The effective fields as functions of applied currents under $H_x = \pm 5000$ Oe. (d) $H_z^{eff} / I_{dc}$ as functions of $H_x$ for WTe$_2$(10)/CoTb(6)/Ta(2) and W(4)/CoTb(6)/Ta(2) (control) heterostructures.

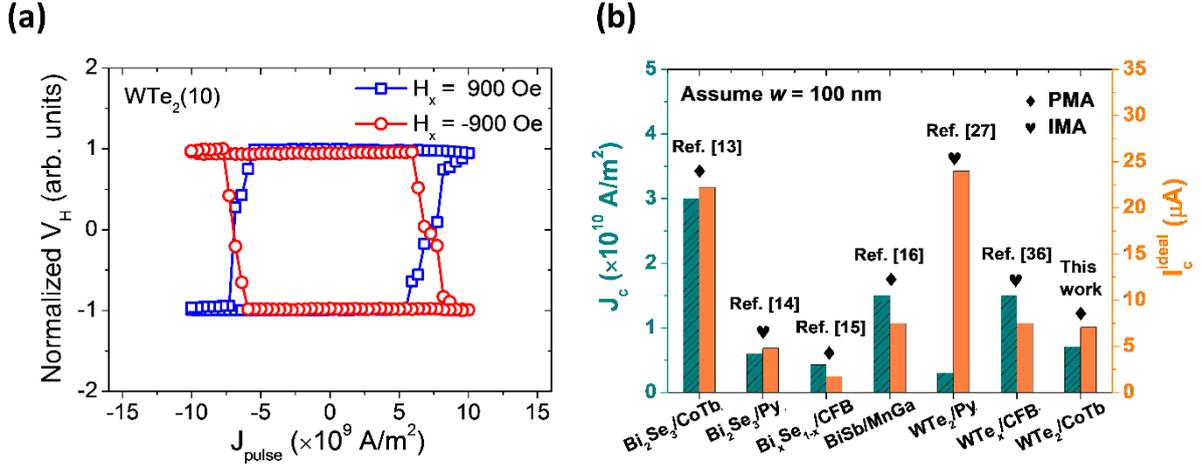

**FIG. 4.** Current-induced magnetization switching in WTe$_2$(10)/CoTb(6)/Ta(2) device. (a) Current-induced magnetization switching from a WTe$_2$(10)/CoTb(6)/Ta(2) device with applied bias fields $H_x = \pm 900$ Oe. (b) Comparison of critical switching current density $J_c$ (left y axis) and ideal critical switching current $I_c^{ideal}$ (right y axis) among TIs and WTe$_2$ based heterostructures.

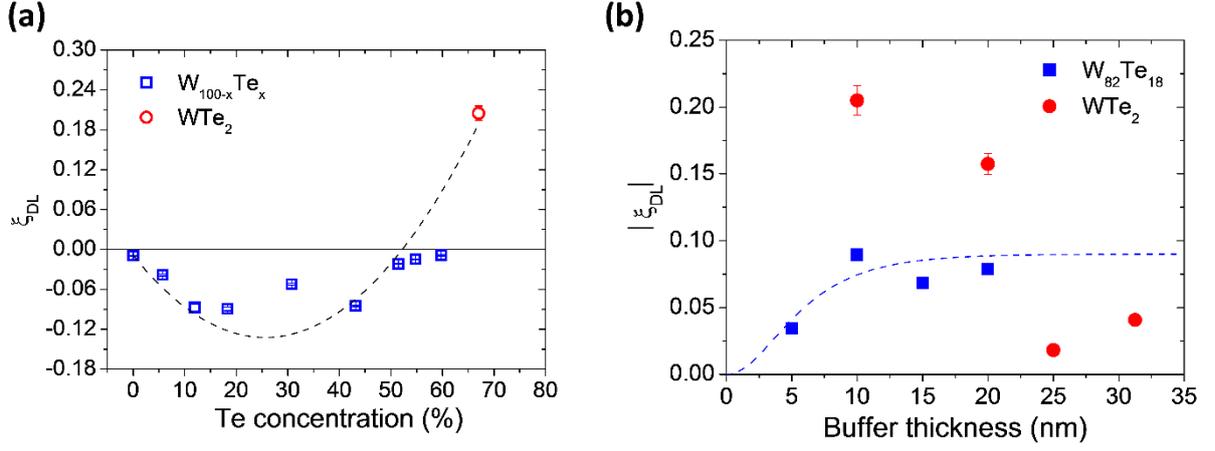

**FIG. 5.** Comparison of SOT from WTe$_2$ and co-sputtered W$_{100-x}$Te$_x$. (a) The DL-SOT efficiency $\xi_{DL}$ of W$_{100-x}$Te$_x$/CoTb heterostructures as a function of Te concentration. The blue open squares represent the devices with co-sputtered W$_{100-x}$Te$_x$, and the red open circle represents the stoichiometric WTe$_2$ case. The dashed line serves as guide to the eye. (b) Buffer layer thickness dependence of $|\xi_{DL}|$ for W$_{82}$Te$_{18}$-based and WTe$_2$-based devices. The blue dashed line represents the fitting of the $|\xi_{DL}(t_{W_{82}Te_{18}})|$ data to a spin diffusion model with spin diffusion length $\lambda_s^{W_{82}Te_{18}} \approx 4.1$ nm.